\documentclass[aps,amsmath,amssymb,showpacs,pre]{revtex4}
\usepackage{graphicx,color}
 \graphicspath{{./}{./figures/}}

\newcommand{\D}{{\rm d}}

\begin{document}
\title{Long time scaling behaviour for diffusion
with resetting and memory}

\author{Denis Boyer}
\email{boyer@fisica.unam.mx}
\affiliation{Instituto de F\'isica, Universidad Nacional Aut\'onoma de M\'exico, D.F. 04510, M\'exico}
\affiliation{Centro de Ciencias de la Complejidad, Universidad Nacional 
Aut\'onoma de M\'exico, D.F. 04510, M\'exico}

\author{Martin R. Evans}
\email{mevans@staffmail.ed.ac.uk}
\affiliation{SUPA, School of Physics and Astronomy,
 University of Edinburgh, 
Peter
Guthrie Tait Road, Edinburgh EH9 3FD, United Kingdom}

\author{Satya N. Majumdar}
\email{majumdar@lptms.u-psud.fr}
\affiliation{Laboratoire de Physique Th\'{e}orique et Mod\`{e}les Statistiques,
 UMR 8626, Universit\'{e}
Paris Sud 11 and CNRS, B\^{a}t. 100, Orsay F-91405, France}

\begin{abstract}
We consider a continuous-space and continuous-time 
diffusion process under  resetting  with memory.
A particle resets to a position chosen from its trajectory in the past
according to a memory kernel.
Depending on the form of the memory kernel,  we show analytically
how different asymptotic behaviours of the variance of  the particle position emerge  at long times. These range from standard  diffusive ($\sigma^2\sim t$) all the way to anomalous ultraslow growth $\sigma^2 \sim \ln \ln t$.
\end{abstract}

\maketitle

\section{Introduction}

Diffusion is a fundamental dynamical process, introduced originally to
describe the motion of molecules and Brownian motion \cite{C43}
and subsequently used to  model  a whole range of stochastic phenomena
from biology to computer science \cite{B93,G06}.  The molecular
origin of diffusion is the random walk of the constituent particles
and one obtains the diffusion equation as a continuum limit of the
random walk. The solution of the diffusion equation leads to the
Gaussian distribution with variance growing linearly in time $\sigma^2(t) \sim t$ which we refer to as diffusive growth.  In  an  
infinite system the Gaussian distribution keeps broadening in time,
whereas in a finite system an equilibrium stationary state  is attained
at long times.

While diffusive processes are abundant in nature, many systems exhibit
anomalous behaviour $\sigma^2(t) \sim t^{\beta_1}$ where $\beta_1 \neq
1$. For example, $\beta_1 <1 $ corresponds to subdiffusion whereas
$\beta_1 > 1 $ corresponds to superdiffusion \cite{BG,KM}.  Models
such as the continuous time random walk (CTRW) provides a microscopic
model for subdiffusive behaviour \cite{MW}. In contrast L\'evy flights or walks
lead to superdiffusive behaviour \cite{BG,KM}.  Another anomalous
behaviour, outside the regime of these models, is the very slow
logarithmic growth with time $\sigma^2 \sim (\ln t)^{\beta_2}$ where
$\beta_2 >0$.  Such logarithmically slow dynamics is usually observed
in glassy or disordered systems. For example in the Sinai model---a
single random walker diffusing in a Brownian random potential---the
variance of the position of the particle grows in time as $\sigma^2
\sim (\ln t)^4$ \cite{BG}.
A natural question is whether microscopic dynamics can generate such
anomalously slow growth of the variance of a single  particle without the explicit presence of disorder or external potential.

Recent studies have  revealed  that a variety of subdiffusive behaviour,
including logarithmic growth can be achieved in a single particle model
with memory-dependent dynamics. In this simple scenario the particle, at a given time, jumps a random distance which may depend on the history 
of the process; these processes are thus typically non-Markovian in nature.
Examples include `Elephant Walk' \cite{ST04},
 `Cookie Random Walks' \cite{Z05,BS08}, 
 `Reinforced Walks' \cite{Davis90,FGP09}, etc.
In the ecological context, a model \cite{GM05,GM06} of animal mobility has been proposed which
incorporates the tendency for an animal to revisit previously
visited sites, a possible mechanism that can explain the formation of home ranges.  A particular case of this model is the
one where, in addition to nearest neighbour random walk dynamics,  
the particle revisits a site  with a probability proportional
to the number of times this site has been visited in the past \cite{BS2014}. 
This model turns out to be exactly solvable \cite{BS2014} and the particle position distribution converges to a Gaussian at long times with a variance growing logarithmically as $\sigma^2 \sim \ln t$.
The model was generalised in \cite{BR2014} to relocation to a site visited in the past, selected according to a  time-weighted distribution with a specific
form. This has been further generalised
to memory kernels that lead to an asymptotic  L\'evy distribution
with index $0< \mu \leq 2$
for the particle position with a length scale growing as 
$(\ln t)^{1/\mu}$ \cite{BP2016}. 
Another simple model with memory that has attracted recent interest is
diffusion with stochastic resetting \cite{MZ99,EM11}. In this case the
dynamics occurs in continuous space-time and, in addition to
diffusion, relocation occurs to the initial position with a constant
rate.  In this case at long times the particle position distribution
converges to a non-Gaussian stationary distribution and the variance
approaches a constant at late times \cite{EM11,MSS15b}.
Several non-Markovian  variations of this  model have been studied recently
\cite{MC16,EM16,PKE16,NG16,MSS15a}.

These studies demonstrate how memory can strongly affect the late time
behaviour of a diffusive process and generate a range of time
dependent growth of the variance of the particle position.  It is then
natural to ask whether a simple model can incorporate all the
different behaviours, found in the various contexts mentioned above,
within a single unified setting.  The purpose of this paper is to
present and analyse such a generalized model.

In the present work we consider diffusion (in continuous space and
time) with a resetting process that relocates the diffusive particle
to a position from its past history.  The model allows a smooth
interpolation between diffusion with resetting to the initial position
and diffusion with time-weighted relocation to positions visited in
the past.  We demonstrate conditions required for the diffusive
particle to attain a localised, nonequilibrium stationary
state. Furthermore, we demonstrate that when the conditions for a
stationary state do not hold, a wider variety of long-time scaling
behaviours than previously encountered are possible: the variance of
the distribution may grow logarithmically in time, subdiffusively as
$t^\beta$ with $\beta< 1$ or diffusively as $t$ with a modified or
unmodified diffusion constant. In addition  we
find for a special case, at the localised-nonlocalised transition, an ultra-slow $\ln\ln t$ growth of the
variance. We note that a similar ultraslow spread of a spatial
distribution has been reported from data on human mobility \cite{SKWB10}.

\section{The Model}
\begin{figure}%[ht]
  \centerline{\includegraphics*[angle=-90,width=\textwidth]{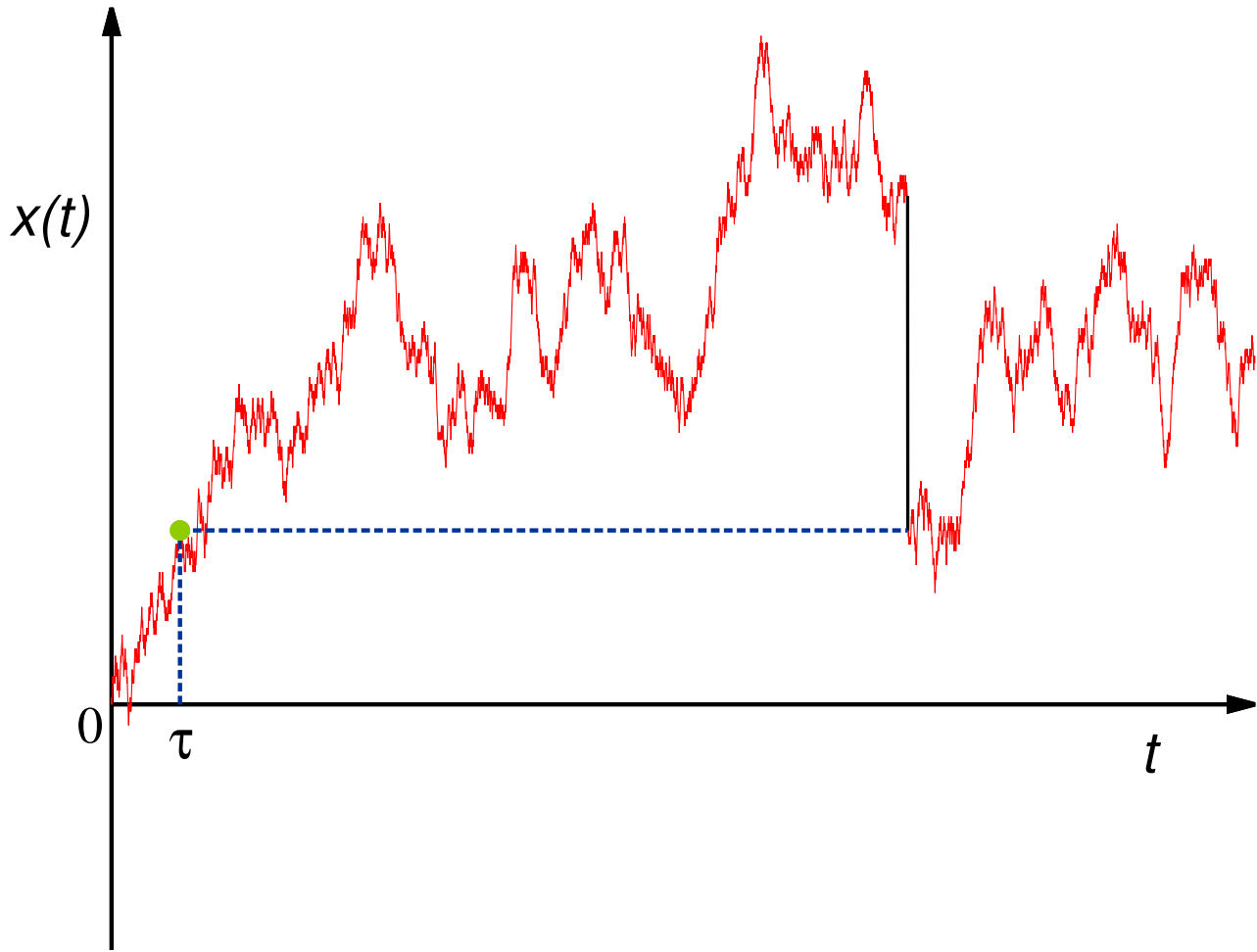}}
  \caption{(Color online) Example of a trajectory starting at the origin
and following the rules of the model considered in this study. 
At fixed rate $r$, the normally diffusive particle resets to a previous 
position $x(\tau)$ where the time $\tau$ is a random variable drawn from a 
given distribution $K(\tau,t)$.}
\label{fig1}
\end{figure}

We consider a continuous-space and continuous-time implementation of
diffusion under resetting with memory. We begin by considering spatial dimension $d=1$ but our results are easily generalisable to arbitrary $d$ (see Section \ref{sec:disc}). We define the resetting rate to
be $r$ and upon resetting at time $t$ the diffusive particle selects a
position $x(\tau)$ from its trajectory, where $\tau$ is chosen with probability density
$K(\tau,t)$. We refer to $K(\tau,t)$ as the memory kernel, which can be any non-negative function and has the
property that $\int_0^t \D \tau K(\tau,t)=1$ for all $t$. This property guarantees
the conservation of total probability. Figure \ref{fig1} illustrates these rules with 
a diffusive trajectory undergoing a first resetting event to a position occupied at a 
former time.

The Master equation for the process reduces to the Fokker-Planck equation
for the probability distribution $p(x,t)$
\begin{equation}
\frac{\partial p(x,t)}{\partial t}= D\, \frac{\partial^2 p(x,t)}{\partial x^2} - r\, p(x,t) + r\,\int_0^t K(\tau,t)\, p(x,\tau)\, d\tau 
\label{fp.1}
\end{equation}
starting from the initial condition $p(x,0)= \delta(x)$. 
The first term on the right hand side of (\ref{fp.1})
represents diffusion with diffusion constant $D$ ; the second term 
represents loss of probability from $x$ due to resetting; the third term 
represents gain of probability into $x$ by choosing a
time $\tau$ in the past with probability density $K(\tau,t)$
and relocating to $x$ with probability density $p(x, \tau)$.
Indeed, by integrating  Eq. (\ref{fp.1}) over $x$ 
and using $\int_0^t \D \tau K(\tau,t)=1$,   it is easy to see that the Master equation conserves the
total probability $ \int {\rm d}x\, p(x,t)=1$ since the integral of the second term is $-r$ and the integral of the third term is $+r$.
Equation (\ref{fp.1}) may be derived following a similar approach to that
used for the discrete time random walk case \cite{BR2014}.

Let us first consider two limiting cases of the kernel $K(\tau,t)$:\\

\noindent (i) {\bf {Case I:}} $K(\tau,t)= \delta(\tau)$. In this case, Eq. (\ref{fp.1}),
upon using $p(x,0)=\delta(x)$, reduces to
\begin{equation}
\frac{\partial p(x,t)}{\partial t}= D\, \frac{\partial^2 p(x,t)}{\partial x^2} - r\, p(x,t) + r\, \delta(x) .
\label{em.1}
\end{equation}
This then corresponds to resetting the walker to its initial position $x_0=0$ with
rate $r$. Hence in this case, the model reduces to the 
`diffusion with stochastic resetting' model studied by Evans and Majumdar~\cite{EM11}. For that model the probability distribution reaches
a localised,  stationary state exponentially decreasing with distance from the resetting position
$\displaystyle p_{\rm st}(x)= \frac{\alpha_0}{2}\,\exp[-\alpha_0\,|x|]$  with
$\alpha_0 = (r/D)^{1/2}$ \cite{EM11}. \\

\noindent (ii) {\bf {Case II:}} $K(\tau,t)= 1/t$. In this case, the Master equation becomes
\begin{equation}
\frac{\partial p(x,t)}{\partial t}= D\, \frac{\partial^2 p(x,t)}{\partial x^2} -r\, p(x,t)+ \frac{r}{t}\int_0^t p(x,\tau)\,d\tau\;.
\label{bs.1}
\end{equation}
This is just the continuous-space and continuous-time version of the discrete-time lattice model  studied by Boyer and Solis-Salas \cite{BS2014}, where the walker
resets with rate $r$ to a previously visited site chosen randomly from all past times.
This is equivalent to saying that a previously visited site is chosen for relocation with a
probability proportional to the number of visits to that site in the past.
In that work it was shown that the late time behaviour is
Gaussian but with a variance growing very slowly (logarithmically) with 
time~\cite{BS2014}.
We will show in Section \ref{sec:uniform} that, as in the corresponding random walk problem, the late time behaviour of Eq. (\ref{bs.1}) is Gaussian (with zero mean): $p(x,t) \sim \exp[-x^2/{2 \sigma^2(t)}]$ 
with a variance growing very slowly (logarithmically) with 
time~\cite{BS2014}: 
$\sigma^2(t)\sim \ln t$.

Our interest is in establishing what further behaviours are possible under resetting  with memory.
To this end, we  choose a  family of memory kernels  $K(\tau,t)$  which allows  a  smooth
interpolation  between these two limiting cases:
\begin{equation}
K(\tau,t)= \frac{\phi(\tau)}{\int_0^t \phi(\tau)\,{\rm d}\tau}
\label{kernel_1}
\end{equation}
where $\phi(\tau)$ can be any non-negative function. 
Thus $K(\tau,t)$ only depends on the present time $t$ through the 
denominator in (\ref{kernel_1}).
By choosing $\phi(\tau)=\delta(\tau)$ one
obtains  Case I, where $p(x,t)$ approaches a non-Gaussian stationary state at late times. In contrast, by choosing
$\phi(\tau)=1$ one recovers case II where the distribution is a Gaussian at late times
with variance growing logarithmically with time $t$. We would like to understand generally how the late time
distribution depends on the choice of $\phi(\tau)$.

We note that  another class of  memory kernel, 
\begin{equation}
K(\tau,t) = \frac{\psi(t-\tau)}{\int_0^t \psi(t-\tau) d\tau}
\label{kernel_br}
\end{equation}  
was considered in the context of random walks with relocation and the special case $\psi(t-\tau)= (1+ t- \tau)^{-\beta}$ (memory decaying as a power law with exponent $\beta>0$) was analysed in detail~\cite{BR2014}. 
Except in the interval $1<\beta<2$, this choice
led to late time Gaussian distributions $p(x,t)\sim \exp[-x^2/{2 \sigma^2(t)}]$, but now the temporal growth of the variance
$\sigma^2(t)$ depends on the choice of the exponent $\beta$. It was found that $\sigma^2(t)\sim t$ for $\beta>2$ and $\sigma^2(t)\sim \ln t$ for $\beta<1$~\cite{BR2014}. 
In the intermediate case $1<\beta<2$, $p(x,t)$ is non-Gaussian and known only through its moments, and $\sigma^2(t)\sim t^{\beta-1}$.

%However, notice that with the choice of the form of the memory kernel in Eq. (\ref{kernel_br}), 
%while one can access Case II (by choosing $\psi(t-\tau)={\rm const.}$), one can not easily access Case I
%by choosing an appropriate $\psi(t-\tau)$. Thus  Case I, namely %$K(\tau,t)=\delta(\tau)$, is not easily included in the
%class of kernels $K(\tau,t) = \psi(t-\tau)/C(t)$. 

The Fokker-Planck equation (\ref{fp.1}) is linear in $p(x,t)$, and hence it is natural to consider the
Fourier transform of $p(x,t)$
\begin{equation}
{\tilde p}(k,t)= \int_{-\infty}^{\infty} p(x,t)\, e^{ikx}\, dx\, .
\label{ft.1}
\end{equation}
Taking the Fourier transform of Eq. (\ref{fp.1}), with the choice for kernel in (\ref{kernel_1}), yields
\begin{equation}
\frac{\partial {\tilde p}(k,t)}{\partial t}= -(r+D\,k^2)\, {\tilde p}(k,t) + 
r\, \frac{\int_0^t \phi(\tau)\,{\tilde p}(k,\tau)\,d\tau}{\int_0^t \phi(\tau)\,d\tau}\, ,
\label{ftfp.1}
\end{equation}
subject to the initial condition, ${\tilde p}(k,0)= 1$. 
In the next two sections we shall consider cases where
(\ref{ftfp.1}) can be solved exactly.
Then in Section ~\ref{sec:gen} we shall  extract the large time
behaviour of the solution in the general case.

\section{ Exact solution for the case $\phi(\tau)= \lambda\, \exp[-\lambda\, \tau]$}\label{sec:exp}

Let us first consider the special case when
\begin{equation}
\phi(\tau)= \lambda\, \exp[-\lambda\, \tau]\, .
\label{exp.1}
\end{equation}
When $\lambda\to \infty$, $\phi(\tau)\to \delta(\tau)$ and we recover Case I. In the opposite limit $\lambda\to 0$, $K(\tau,t)=1/t$ and we recover Case II. 
However as the limits $\lambda \to 0$ and $t\to \infty$ do not commute
we shall consider the case $\lambda =0$ separately in the next section. 

Substituting $\phi(\tau)=\lambda\, 
\exp[-\lambda\, \tau]$ in Eq. (\ref{ftfp.1}) gives
\begin{equation}
\left(1-e^{-\lambda t}\right)\, \left[\frac{\partial {\tilde 
p}(k,t)}{\partial t}+(r+D\,k^2)\,{\tilde p}(k,t)\right]= 
r\lambda\, \int_0^t {\tilde p}(k,\tau)\,e^{-\lambda \tau}\, d\tau\, .
\label{exp.2}
\end{equation}
Differentiating once more with respect to $t$ reduces 
(\ref{exp.2})  to a second order differential equation
\begin{equation}
\left(e^{\lambda t}-1\right)\left[\frac{\partial^2 {\tilde 
p}(k,t)}{\partial t^2}+(r+D\,k^2)\, 
\frac{\partial {\tilde p}(k,t)}{\partial t}\right]+ 
\lambda\left[\frac{\partial {\tilde p}(k,t)}{\partial t}+D\,k^2 {\tilde 
p}(k,t)\right]=0\, .
\label{exp.3}
\end{equation}
To solve this equation, we first make a change of variable, from $t$ to
$y= e^{-\lambda t}$. Then, 
\begin{equation}
{\tilde p}(k,t)= W(y={\rm e}^{-\lambda t})
\label{pwy}
\end{equation}
where $W(y)$ depends implicitly on $k$. Substituting 
(\ref{pwy})
in Eq. (\ref{exp.3}), we get
after a few steps of straightforward algebra, a second order differential 
equation for $W(y)$
\begin{equation}
y(1-y)\frac{d^2W(y)}{dy^2} + \left[1- 
\frac{1}{\lambda}(r+D\,k^2)-\left(2-\frac{r+D\,k^2}{\lambda}\right)\,y\right] 
\frac{dW(y)}{dy}+ \frac{D\,k^2}{\lambda}\, W(y)=0\, .
\label{wdiff.1}
\end{equation}
As $0\le e^{-\lambda t} \le 1$,  the second order differential equation (\ref{wdiff.1}) holds in the regime $0\le y\le 1$. The initial condition, ${\tilde p}(k,t=0)=1$ and the fact that $y=e^{-\lambda t}\to 1$ as $t\to 0$, translates 
into one boundary condition for $W(y)$
\begin{equation}
W(y=1) =1\, .
\label{bc.1}
\end{equation}
To solve a second order equation, we generally need two boundary conditions. However, we note that the original equation (\ref{exp.2})
(before taking the derivative with respect to time) is a first order differential equation. Hence, it turns out that just one initial
condition in $t$ (or equivalently one boundary condition  at $y=1$,
$W(1)=1$) is enough to fix the full solution of
(\ref{wdiff.1}).

Fortunately, Eq. (\ref{wdiff.1}) has the form of a standard hypergeometric differential equation
\begin{equation}
y(1-y) \frac{d^2W(y)}{dy^2} + \left[c-(a+b+1)y\right]\frac{dW(y)}{dy}- ab W(y)=0\, ,
\label{hyper.1}
\end{equation}
once we identify
\begin{eqnarray}
a& =& \frac{1}{2}\left[ c + \sqrt{c^2+ \frac{4D\,k^2}{\lambda}}\right] 
\label{a.1} \\
b& =& \frac{1}{2}\left[ c - \sqrt{c^2+ \frac{4D\,k^2}{\lambda}}\right] 
\label{b.1} \\
c &=& 1- \frac{r+D\,k^2}{\lambda}= a+b\;. \label{c.1} 
\end{eqnarray}
The general solution of the hypergeometric differential equation (\ref{hyper.1}) can be
written as a linear combination~\cite{AS}
\begin{equation}
W(y)= A\, F(a,b,c\,;\, y) + B\, y^{1-c}\, F(a-c+1,b-c+1,2-c\,;\,y)
\quad\quad 0\le y\le 1\, ,
\label{hypersol.1}
\end{equation}
where $F(a,b,c\,;\,y)$ is the standard hypergeometric function and $A$ and 
$B$ are two arbitrary constants, yet to be determined. To fix the two 
constants $A$ and $B$, we will use the boundary condition $W(y=1)=1$ in 
Eq. (\ref{bc.1}). In order to do so, it is convenient to
use the following identity~\cite{AS}
\begin{equation}
F(a,b,a+b\,;\,y)= 
\frac{\Gamma(a+b)}{\Gamma(a)\Gamma(b)}\,\sum_{n=0}^{\infty} 
\frac{(a)_n\,(b)_n}{(n!)^2}\,
\left[2\psi(n+1)-\psi(a+n)-\psi(b+n)-\ln(1-y)\right]\,(1-y)^n
\label{hypergeo.iden}
\end{equation}
where $(a)_n= a(a+1)(a+2)\ldots (a+n-1)$ is the Pochhammer symbol
and $\psi(z)= \Gamma'(z)/\Gamma(z)$. As $y\to 1$, keeping only the
leading $n=0$ term gives
\begin{equation}
F(a,b,a+b\,;\,y\to 1)\to \frac{\Gamma(a+b)}{\Gamma(a)\Gamma(b)}\,
\left[2\psi(1)-\psi(a)-\psi(b)-\ln(1-y)\right]\,.
\label{hypergeo.y1}
\end{equation}
Using this result in Eq. (\ref{hypersol.1}) gives
\begin{eqnarray}
W(y\to 1)&=& A\, \frac{\Gamma(a+b)}{\Gamma(a)\Gamma(b)}\,
\left[2\psi(1)-\psi(a)-\psi(b)-\ln(1-y)\right]\nonumber\\
&& + B\, \frac{\Gamma(2-a-b)}{\Gamma(1-a)\Gamma(1-b)}\,
\left[2\psi(1)-\psi(1-a)-\psi(1-b)-\ln(1-y)\right]\, .
\label{wy1}
\end{eqnarray}
In order to enforce the boundary condition $W(y=1)=1$
the coefficient of the leading diverging term $\ln(1-y)$ in Eq. 
(\ref{wy1}) must vanish. In addition, the subleading constant term
must be equal to unity. This then provides two equations to determine
the two unknown constants $A$ and $B$. These two equations read
\begin{eqnarray}
& A\, \frac{\Gamma(a+b)}{\Gamma(a)\Gamma(b)}+ B\, 
\frac{\Gamma(2-a-b)}{\Gamma(1-a)\Gamma(1-b)}=0 \label{eq1} \\
& A\, \frac{\Gamma(a+b)}{\Gamma(a)\Gamma(b)}\,
\left[2\psi(1)-\psi(a)-\psi(b)\right]
+B\, \frac{\Gamma(2-a-b)}{\Gamma(1-a)\Gamma(1-b)}\,
\left[2\psi(1)-\psi(1-a)-\psi(1-b)\right]=1 \, . \label{eq2}
\end{eqnarray} 
Using (\ref{eq1}) in (\ref{eq2}) gives
\begin{equation}
A\, 
\frac{\Gamma(a+b)}{\Gamma(a)\Gamma(b)}\,
\left[\psi(1-a)+\psi(1-b)-\psi(a)-\psi(b)\right]=1\, .
\label{a1}
\end{equation}
This determines $A$. The other constant $B$ can be
determined using (\ref{eq1}). The expressions for $A$ and $B$ can be 
further simplified by
using the relation~\cite{AS}, $\psi(1-a)-\psi(a)=\pi\, \cot(\pi 
a)$ and the two identities:  $\cot(\pi a)+\cot(\pi b)= 
\sin(\pi(a+b))/[\sin(\pi a)\sin(\pi b)]$ and  $\Gamma(a)\Gamma(1-a)= 
\pi/\sin(\pi a)$
which eventually lead to
\begin{eqnarray}
A &=& \frac{\Gamma(1-a-b)}{\Gamma(1-a)\Gamma(1-b)} \label{a2} \\
B &=& - \frac{1}{1-a-b}\, \frac{\Gamma(a+b)}{\Gamma(a)\Gamma(b)}\, . 
\label{b2}
\end{eqnarray} 
Plugging these expressions into (\ref{wy1}) and equating
$W(y) ={\tilde p}(k,t)$
gives the
exact time-dependent solution at all times $t$
\begin{eqnarray}
{\tilde p}(k,t)&=& 
%W\left(e^{-\lambda t}\right)= 
\frac{\Gamma(1-a-b)}{\Gamma(1-a)\Gamma(1-b)}\,F(a,b,a+b;\, e^{-\lambda t})\nonumber \\
&&- \frac{1}{1-a-b}\, \frac{\Gamma(a+b)}{\Gamma(a)\Gamma(b)}\, 
e^{-\lambda(1-c)t}\, F(1-b,1-a,2-a-b;\, e^{-\lambda t})\, ,
\label{final_sol1}
\end{eqnarray}
where, we recall that $a$ and $b$ are given in Eqs. (\ref{a.1}) and 
(\ref{b.1}).

It is evident from this exact solution in Eq. (\ref{final_sol1}) that
for the choice, $\phi(\tau)=\lambda \exp[-\lambda \tau]$ with $\lambda>0$, the
system always reaches a stationary state as $t\to \infty$. Taking $t\to \infty$
in Eq. (\ref{final_sol1}), using $F(a,b,c;\, y=0)=1$ and $c<1$, one obtains 
exactly the stationary probability distribution (in  Fourier space)
\begin{equation}
{\tilde p}_{\rm st}(k)= \frac{\Gamma(1-a-b)}{\Gamma(1-a)\Gamma(1-b)}\, ,
\label{stat_sol.1}
\end{equation}
with $a$, $b$ given in Eqs. (\ref{a.1}) and (\ref{b.1}).
We can rewrite expression (\ref{stat_sol.1}) as
\begin{equation}
{\tilde p}_{\rm st}(k)= B(1-a-b, a) \frac{\sin \pi a}{\pi}
\label{stat_sol.2}
\end{equation}
where $B(x,y)$  is the usual Beta function \cite{AS}.

Let us first verify that in the limit $\lambda\to \infty$ limit, we recover
the results of Ref.~\cite{EM11}. As $\lambda\to \infty$, we get from
Eqs. (\ref{a.1}), (\ref{b.1}) and (\ref{c.1})
\begin{eqnarray}
c &= &  1- \frac{r+D\,k^2}{\lambda} \label{c.2} \\
a & \to & 1- \frac{r}{\lambda} \label{a.2} \\
b & \to & - \frac{D\, k^2}{\lambda} \, . \label{b.2} 
\end{eqnarray}
Using $\Gamma(x)\to 1/x$ as $x\to 0$ in (\ref{stat_sol.1}), gives
\begin{equation}
{\tilde p}_{\rm st}(k)\xrightarrow[\lambda\to \infty]{} \frac{r}{r+ D\, k^2}\,.
\label{stat_sol.3}
\end{equation}
Inverting the Fourier transform exactly reproduces the stationary solution in
real space obtained in Ref.~\cite{EM11}
\begin{equation}
p_{\rm st}(x) =\sqrt{\frac{r}{4D}}\,\exp\left[-\sqrt{\frac{r}{D}}\,|x|\right]\, .
\label{stat_sol.em}
\end{equation}

For finite $\lambda$, one notices that the right hand side
of Eq. (\ref{stat_sol.1}) (or equivalently  Eq. (\ref{stat_sol.2}))  has poles when $1-a-b=1-c= 1- (r+D\,k^2)/\lambda= -m$
where $m=0,1,2\ldots$. In other words, the poles in the complex $k$ plane occur at
$k= \pm i \sqrt{(r+\lambda m)/D}$. Hence, the solution
in the real space can be written as a sum over exponentials (each coming from
a pole in the lower half of the complex $k$-plane for $x>0$ and from
the upper half if $x<0$)
\begin{equation}
p_{\rm st}(x) = \sum_{m=0}^{\infty} A_m e^{-\sqrt{\frac{r+\lambda m}{D}}\, |x|}\, ,
\label{stat_sol_gen}
\end{equation}
Thus at finite $\lambda>0$ the leading term of $p_{st}(x)$ at large $|x|$ is still given by $e^{-\sqrt{r/D}|x|}$. The coefficients $A_m$'s can be computed from the residues at the relevant poles
and one obtains
\begin{equation}
A_m = \frac{\lambda}{2\pi D^{1/2}} \frac{(1-a_m)_m}{m! (r+m\lambda)^{1/2} }
\sin \pi a_m
\end{equation}
where
\begin{equation}
a_m = \frac{1}{2}\left[ 1+m + \left((1-m)^2-4r/\lambda\right)^{1/2}\right]\;.
\end{equation}

\section{Exact solution for uniform memory kernel: $\phi(\tau)= 1$}
\label{sec:uniform}
In the previous section we considered $\phi(\tau)$ given by (\ref{exp.1})
for $\lambda >0$.
Since the asymptotic results  only hold for $\lambda >0$,
we need to consider the special case $\lambda\to 0$ (which corresponds to
$\phi(\tau)=1$) separately, as presented below.
This case  actually corresponds to the continuous space-time  limit of \cite{BS2014}.

We consider $\phi(\tau)= 1$ or $K(t,\tau) = 1/t$
in which case  we obtain from (\ref{fp.1})
\begin{equation}
t \frac{\partial^{2} \tilde{p}(k,t)}{\partial t^2} + (1+(r+Dk^2)t) \frac{\partial \tilde{p}(k,t)}{\partial t} + Dk^2\tilde{p}(k,t)=0\, ,
\label{ft1}
\end{equation}
an equation that can be also deduced from the $\lambda \to 0$ limit of (\ref{exp.3}).

We now define
 $u=Dk^2+r$ and $v=Dk^2$ in (\ref{ft1}). Then
\[t \frac{\partial^{2} \tilde{p}(k,t)}{\partial t^2} + (1+ut) \frac{\partial \tilde{p}(k,t)}{\partial t} + v\tilde{p}(k,t)=0.\]
Making a further substitution $z=-ut$ gives
\begin{equation}
z\frac{\partial^{2} \tilde{p}(k,z)}{\partial z^2} + (1-z) \frac{\partial \tilde{p}(k,z)}{\partial z} -\frac{v}{u}\tilde{p}(k,z)=0
\label{ft2}
\end{equation}
which is Kummer's equation 
\begin{equation}
z\frac{\partial^{2} w(z)}{\partial z^2} + (b-z) \frac{\partial w(z)}{\partial z} -aw(z)=0
\end{equation}
with $b=1$, $a=\frac{v}{u}$.
We note that Kummer's equation (\ref{ft2}) is a degenerate case of the
hypergeometric equation (\ref{hyper.1}) and the Kummer functions are confluent hypergeometric functions.
For  all values of $a$ and $z$, and $b$ not a non-positive integer,  the Kummer functions are given by
\begin{equation}
M(a,b,z)=\sum_{n=0}^\infty \frac{(a)_{n}z^n}{(b)_{n}n!}\;,
\end{equation}
where $(a)_n$ is again the Pochhammer symbol.
As the other solution to (\ref{ft2}) diverges we deduce
$\tilde{p}(k,t) = C(k)M \bigg(\frac{v}{u},1,-ut\bigg)$ where $C(k)$ is to be fixed. From the initial condition $x_0 =0$ and the fact that in the limit $z \to 0, M(a,b,z)\to 1$ 
we find $C(k)=1$.

In principle the exact probability density is obtained by inverting the inverse Fourier transform,
\begin{equation}
p(x,t)=\frac{1}{2\pi}\int_{-\infty}^\infty dk \, e^{-ikx} \,
\sum_{n=0}^\infty 
 \left(\frac{Dk^2}{Dk^2+r}\right)_n \frac{(-(Dk^2+r)t)^n}{n!\, n!}
\label{puni}
\end{equation}
but  the inversion is difficult to carry out.

However, we can easily compute the variance
of the distribution (mean squared displacement),
exact for all times, from the Fourier transform using the formula
\begin{equation}
\left<x^2\right> =
-\left.\frac{\partial^2 \tilde p(k,t)}{\partial k^2}\right|_{k=0}\;.
\end{equation}
Now  we find
\begin{eqnarray}
\left. \frac{\partial^2}{\partial k^2}(a)_n \right|_{k=0}
=
\left. \frac{\partial^2}{\partial k^2}
\left[\left (\frac{Dk^2}{Dk^2+r}\right)_n\right] \right|_{k=0}
= \frac{2D}{r}(n-1)!\;,
\end{eqnarray}
so that we obtain
\begin{eqnarray}
\left<x^2\right> 
&=&\frac{2D}{r}\sum_{n=1}^{\infty}\frac{(rt)^n(-1)^{n+1}}{n!n}\;.
\label{x21}
\end{eqnarray}
The asymptotic, large $t$ form of the second moment can be obtained
by noting 
\begin{eqnarray}
\left<x^2\right>
&=&\frac{2D}{r}\left[E_1(rt)+\gamma+\ln (rt)\right],
\label{x22}
\end{eqnarray}
where  the exponential integral $E_1(z)$ is given by
\begin{equation}
E_1(z) =\int_z^{\infty}dx\frac{e^{-x}}{x}
=-\gamma-\ln z -\sum_{n=1}^{\infty}\frac{(-1)^{n}z^n}{n!n}
\end{equation}
 and $\gamma=0.5772\ldots$ is Euler's constant. Substituting the asymptotic expansion of $E_1(z)$ 
\begin{equation}E_1(z)\sim\frac{e^{-z}}{z}\left(1-O\left(\frac{1}{z}\right)\right)
\qquad\text{for}\qquad z\to\infty,\end{equation} 
into (\ref{x22}),
we obtain
\begin{equation}
\left<x^2\right>=\frac{2D}{r}\left(\ln(rt)+\gamma+\frac{e^{-rt}}{rt}+\ldots\right) \quad \text{for}\qquad t\gg 1.
\end{equation}
This result is consistent  with the that of ~\cite{BR2014}
for the discrete, random walk case.

\section{General memory kernel}
\label{sec:gen}
We now consider a general memory kernel given by a general function $\phi(\tau)$. It is possible
to  analyse Eq. (\ref{ftfp.1}) at late times 
$t$  according to the large $\tau$ behaviour of   
$\phi(\tau)$.  We find a rather rich variety of
asymptotic behaviour of $\sigma^2(t)$ which we summarize below.

\begin{itemize}
\item When $\phi(\tau)$ decays more quickly  than $1/\tau$ for large $\tau$
i.e. $\tau \phi(\tau) \to 0 $ as $\tau\to \infty$
there is a stationary distribution $p_{st}(x)$.

\item When $\phi(\tau)$ increases, or  decays as, or more slowly than, $1/\tau$ for large $\tau$,
i.e., $\tau \phi(\tau) > 0 $ as $\tau\to \infty$,
there is no stationary distribution. Instead there is a late time behaviour
in which the time-dependent distribution takes a Gaussian form with
variance $\sigma^2(t)$.

The time dependence of the variance has different classes of behaviour as follows:
\begin{enumerate}
\item for $\phi(\tau) \sim 1/\tau$, $\sigma^2(t)\sim \ln \ln t$;
\item for $\phi(\tau) \sim \tau^{\alpha}$ with $\alpha > -1$, $\sigma^2(t)\sim \ln t$;
\item for $\phi(\tau) \sim \exp(a \tau^\beta)$ where  $0<\beta <1$ and $a$  is a positive constant, $\sigma^2(t)\sim  t^\beta$;
\item for $\phi(\tau) \sim \exp(a \tau)$  where $a$ is a positive constant, $\sigma^2(t)\simeq  \displaystyle \left(\frac{2a}{a+r}\right) Dt$;
\item for $\phi(\tau) \sim \exp(a \tau^\beta)$ where $\beta > 1$  and $a$ is a positive constant, $\sigma^2(t)\simeq  2Dt$. 
\end{enumerate}
\end{itemize}
We now discuss these different cases separately.

\subsection{$\phi(\tau)$ decaying faster than $1/\tau$ as $\tau\to 
\infty$}

Consider the case where $\phi(\tau)\ll 1/\tau$ as $\tau\to \infty$. In this 
case, the integral $\int_0^t \phi(\tau) d\tau$ appearing in the 
denominator of the second term in Eq. (\ref{ftfp.1}) converges as
$t\to \infty$. Hence, we 
expect that as $t\to \infty$, the system approaches a stationary
state, and the stationary solution (in Fourier space) is obtained by 
setting the rhs of Eq. (\ref{ftfp.1}) to zero
\begin{equation}
(r+ Dk^2) {\tilde p}_{\rm st}(k)=  r \frac{\int_0^\infty \phi(\tau)\,
{\tilde p}(k,\tau)\,d\tau}{\int_0^\infty \phi(\tau)\,d\tau}\, .
\label{stat_1.1}
\end{equation}
Of course, determining ${\tilde p}_{\rm st}(k)$ explicitly requires
knowledge of the full time-dependent solution in the right hand side of
(\ref{stat_1.1}). Nevertheless, it is clear that the integrals in the right hand side of  (\ref{stat_1.1}) converge and that therefore there
is a stationary state solution 
${\tilde p}_{\rm st}(k)$
 as $t\to \infty$.

\subsection{Power-law family: $\phi(\tau)\sim  \tau^{\alpha}$ as $\tau\to 
\infty$ with $\alpha>-1$}\label{sec.alpha}

Let us assume $\phi(\tau)\sim \tau^{\alpha}$ as $\tau\to \infty$
and that $\alpha>-1$.
In this case, to analyse Eq. (\ref{ftfp.1}) at late times, we make the
ansatz 
\begin{equation}
{\tilde p}(k,t) \sim \frac{A}{t^{\gamma}}\quad {\rm with} \quad
0<\gamma<\alpha+1 \, .
\label{ansatz_II.1}
\end{equation}
Substituting this ansatz in Eq. (\ref{ftfp.1}), one finds that the
lhs decays as $t^{-\gamma-1}$ while the rhs decays as $t^{-\gamma}$.
Thus the pre-factor of the term of order $t^{-\gamma}$ in the rhs must be $0$, which fixes the exponent $\gamma$
\begin{equation}
\gamma= (\alpha+1)\frac{D\,k^2}{r+ D k^2}\, ,
\label{gamma_II.1}
\end{equation}
for any $k$. The above expression is consistent with the {\it a priori} assumption 
$\gamma<\alpha+1$. Hence, as $k\to 0$,
\begin{equation}
{\tilde p}(k,t) \sim A t^{-\gamma}=A \exp[- \gamma\, \ln(t)]\sim 
A\exp\left[-\frac{(\alpha+1)D}{r}\, k^2\, \ln(t)\right]\, .
\label{pkt_II.1}
\end{equation}
The normalisation condition ${\tilde p}(k=0,t)=1$ imposes $A\to 1$ as $k\to 0$. Inverting the Fourier transform, one obtains a Gaussian distribution
$p(x,t) \sim \exp[- x^2/2\sigma^2(t)]$ where the variance grows with
time logarithmically
\begin{equation}
\sigma^2(t) \approx \frac{2D}{r}(\alpha+1)\, \ln (t)\, .
\label{pxt_II.1}
\end{equation}
In particular, for $\alpha=0$, one recovers the result of
Section (\ref{sec:uniform}). Note that Gaussian diffusion with logarithmic variance as above is also observed in
certain Markovian processes, for example, for Brownian motion when the diffusion coefficient 
decays as $1/t$ at large times \cite{BCCM15}.

\subsection{The borderline case when $\phi(\tau)\sim 
1/\tau$ as $\tau\to \infty$, i.e, $\alpha=-1$}

In this borderline case, the following ansatz is appropriate
\begin{equation}
{\tilde p}(k,t) \sim \frac{A}{(\ln t)^{\gamma}}
\label{border_II.1}
\end{equation}
where the exponent $\gamma>0$ is yet to be determined. Substituting
this ansatz into the rhs of Eq. (\ref{ftfp.1}) and neglecting the lhs
(which decays faster), one finds that
\begin{equation}
\gamma= \frac{Dk^2}{r+Dk^2}\, .
\label{beta_II.1}
\end{equation}
As above, $A$ must tend to $1$ when $k\to 0$. Hence, for small $k$,
\begin{equation}
{\tilde p}(k,t) \sim \exp[-\beta \ln\ln t]\sim \exp[-\frac{D}{r}k^2 
\ln\ln(t)]\, ,
\label{pkt_border_II.1}
\end{equation}
indicating  a Gaussian in real space with variance growing 
extremely slowly for large $t$ as
\begin{equation}
\sigma^2(t)\approx \frac{2D}{r}\, \ln\ln(t) \, .
\label{variance_border_II.1}
\end{equation} 

\subsection{Stretched exponential family: $\phi(\tau)\sim \exp[a 
\tau^\beta]$ as $\tau\to \infty$ and with $a>0$ and $\beta>0$}
 
Here we consider a class of functions belonging to the family
\begin{equation}
\phi(\tau) \sim  \exp[ a\, \tau^{\beta}] \quad {\rm where}\quad a>0
\label{sexp_III.1}
\end{equation}
where the stretching exponent $\beta$ is positive. In this model recent times 
($\tau$ close to $t$) are thus sampled much more often
than early times $\tau\ll t$. This form of memory decay is controlled with the parameter 
$\beta$, larger values of $\beta$ corresponding to shorter memory ranges. 
We consider below three subclasses: (i) $0<\beta<1$ (ii) $\beta=1$ and (iii) $\beta>1$.

\subsubsection{Case $0<\beta<1${\rm :}}

In this case, the integral $\int_0^t \phi(\tau) d\tau$ behaves for large 
$t$ (to leading order) as
\begin{equation}
\int_0^t \phi(\tau) d\tau \sim \int_0^t e^{a\,\tau^\beta} d\tau\sim 
\frac{1}{a\beta}\, t^{1-\beta}\, e^{a\,t^\beta}\, .
\label{b1.1}
\end{equation}
Now, we make the ansatz that for large $t$
\begin{equation}
{\tilde p}(k,t) \sim A\, \exp[-c\, t^{\beta}]\, 
\label{b1.2}
\end{equation}
where we assume $0<c<a$ (to be justified {\it a posteriori}).
We next substitute this ansatz in Eq. (\ref{ftfp.1}). The
integral in the numerator of the second term on the rhs
of Eq. (\ref{ftfp.1}) can be estimated, to leading order in large $t$,
as (see Eq. (\ref{b1.1}))
\begin{equation}
\int_0^t \phi(\tau) {\tilde p}(k,\tau)\,d\tau\sim A \int_0^t 
e^{(a-c)\,\tau^{\beta}}\, d\tau \sim \frac{A}{(a-c)\beta}\, t^{1-\beta}\, 
e^{(a-c) t^\beta}\, .
\label{num_b1.3}
\end{equation}
Substituting these results in Eq. (\ref{ftfp.1}) gives
\begin{equation}
-c \beta\, t^{\beta-1} e^{-c t^{\beta}}\approx -(r+Dk^2)\, e^{-c 
t^{\beta}} + \frac{r a}{a-c}\, e^{-c t^{\beta}}\, .
\label{b1.4}
\end{equation}
For $\beta<1$, the lhs decays faster than rhs and equating the rhs to $0$
fixes the constant $c$
\begin{equation}
c= a\frac{Dk^2}{r+Dk^2}<a \, 
\label{b1.5}
\end{equation}
which is consistent with the {\it a priori} assumption $c<a$. Hence, this gives, 
for small $k$,
\begin{equation}
{\tilde p}(k,t) \sim  \exp\left[-\frac{aD}{r} k^2\, t^{\beta}\right]\, 
\label{b1.6}
\end{equation}
indicating once again a Gaussian distribution for $p(x,t)$ with variance
growing subdiffusively with time
\begin{equation}
\sigma^2(t) \approx \frac{2 a D}{r}\, t^{\beta}\quad {\rm where}\quad 
0<\beta<1\, .
\label{b1.7}
\end{equation}

\subsubsection{Case $\beta=1${\rm :}}

We now consider the case, $\phi(\tau)\sim e^{a\,\tau}$ as $\tau\to 
\infty$.
In this case, the correct ansatz for ${\tilde p}(k,t)$ turns out to be
\begin{equation}
{\tilde p}(k,t) \sim A\, {\rm e}^{-b\, t}
\label{b2.1}
\end{equation}
where we assume $b<a$ {\it a priori} (to be justified {\it a posteriori}). Substituting
this ansatz in Eq. (\ref{ftfp.1}), we find that now the lhs and 
the rhs both
scale as $e^{-bt}$ for large $t$. Equating the two sides gives the 
identity
\begin{equation}
(r+Dk^2-b)= r\frac{a}{a-b}\,.
\label{b2.2}
\end{equation}
This is a quadratic equation for $b$ which can be solved to give
\begin{equation}
b_{\pm}= \frac{1}{2}\left[r+Dk^2+a\pm \sqrt{(r+Dk^2+a)^2-4aDk^2}\right]\,.
\label{b2.3}
\end{equation}
It can be shown easily that $b_{-}<a$. The condition $b<a$ thus forces us 
to choose $b_{-}$,
\begin{equation}
b= b_{-}= \frac{1}{2}\left[r+Dk^2+a - 
\sqrt{(r+Dk^2+a)^2-4aDk^2}\right]\,. 
\label{b2.4}
\end{equation}
For $k\to 0$, we have
\begin{equation}
b\approx \frac{a}{r+a}\, Dk^2\, .
\label{b2.5}
\end{equation}
Hence, for small $k$ and large $t$,
\begin{equation}
{\tilde p}(k,t) \sim \exp\left[-\frac{aD}{r+a} k^2\, 
t\right]\, .
\label{b2.6}
\end{equation}
Once again, $A\to 1$ at small $k$ due to normalisation.
Inverting the Fourier transform, one again gets a Gaussian distribution 
for $p(x,t)$ with variance growing diffusively for large $t$
\begin{equation}
\sigma^2(t) \approx \frac{2a}{r+a}\,D\, t\, .
\label{b2.7}
\end{equation}
Diffusion is normal in this case, but the effective diffusion coefficient 
of the process, $aD/(r+a)$, is smaller than the bare coefficient $D$ for 
any positive resetting rate $r$.

\subsubsection{Case $\beta>1${\rm :}}

Finally, we consider the class of functions where
$\phi(\tau)\sim  e^{a\,\tau^\beta}$ for large $\tau$ with
the exponent $\beta>1$. In this case, the correct ansatz for ${\tilde 
p}(k,t)$ turns out  to be an exponential, i.e.,
\begin{equation}
{\tilde p}(k,t) \sim A\, e^{-b\, t}
\label{b3.1}
\end{equation}
where the constant $b$ is yet to be determined. Substituting this ansatz
in Eq. (\ref{ftfp.1}), we find that (as in the case $\beta=1$), the lhs 
and rhs both scale as $\exp[-b\, t]$ at late times and equating them
fixes the constant $b$ simply as
\begin{equation}
b_1= D\, k^2\, .
\label{b3.2}
\end{equation}
Note that $b$ is now completely independent of the resetting rate 
$r$. Hence, in this case, one recovers {\em pure} diffusion (as in 
the $r=0$ case),
\begin{equation}
{\tilde p}(k,t) \sim A\, e^{- D\,k^2 t}\, .
\label{b3.3}
\end{equation}
When inverted, one obtains a Gaussian $p(x,t)$ (the standard 
purely diffusive propagator) with variance growing linearly with $t$ 
(independent of $r$)
\begin{equation}
\sigma^2(t) \approx 2\,D\, t\, .
\label{b3.4}
\end{equation}
Thus, for this class of memory kernel $\phi(\tau)$, resetting is not
effective at all, and the walker undergoes normal diffusion asymptotically 
at late times.

\section{Discussion}\label{sec:disc}
In this work we have considered diffusion with resetting to a position 
from the past chosen according to memory kernel (\ref{kernel_1}).
We have demonstrated that a remarkable range of  long time behaviours 
for the variance of the distribution of the diffusive particle are made possible 
through different choices of the function $\phi(\tau)$,
as summarised in Section {\ref{sec:gen}.

We recover previous results for diffusion with resetting \cite{EM11} and
the preferential relocation model \cite{BS2014} as limiting cases.
For functions $\phi(\tau)$ decaying faster than $1/\tau$, the distribution of 
the position converges towards a non-equilibrium stationary state at large times, 
whereas slower decays produce
logarithmic diffusion. At the transition point between these two regimes, for
$\phi(\tau) \sim 1/\tau$, we obtain a Gaussian with variance growing
ultra-slowly as $\ln \ln t$, an unusual diffusive behaviour. 
When the memory function $\phi$ {\it increases} with $\tau$
as a stretched exponential with exponent $\beta<1$ (which mimics a form of memory
decay since recent times are sampled much more frequently than early times) 
non-Markovian effects become weaker but the particle is still subdiffusive and
characterised by a Gaussian distribution with variance $t^{\beta}$. 
This law is replaced by normal diffusion for $\beta\ge1$.

Let us comment that although we have considered a one-dimensional process
the results are easily generalisable to arbitrary spatial dimension.
In the case where a stationary state exists we expect 
the form of the stationary state to depend on dimension as was the case in
\cite{EM14}. However in the 
case of a time-dependent scaling distribution
we expect the scaling distribution to be a $d$ dimensional Gaussian distribution with  width growing in the same way as the one-dimensional case we have presented  here. The reason being that when we generalise to $d$ dimension the  equation (\ref{ftfp.1}) obeyed by the Fourier transform, it 
remains the same but with $k$ replaced by $| \underline k|$, where 
the Fourier transform of $p(\underline  x,t)$ is now 
\begin{equation}
{\tilde p}( \underline k,t)= \int_{-\infty}^{\infty} p(\underline x,t)\, e^{i \underline k\cdot \underline x}\, d^dx\, .
\label{ft.2}
\end{equation}
Therefore the small $k$ expansions we have used in Section \ref{sec:gen}
will also apply in the dimension $d$ case and we will obtain the same time dependences for the width.

We conclude by mentioning that the recurrence properties of 
random walks with memory and resetting are still poorly understood.
Due to the non-Markovian nature of these processes, the study
of their first passage times is a challenging problem \cite{BMS}.\\[2ex]

\noindent {\bf Acknowledgements}
We acknowledge that some results of section IV
were  derived in the Master's Thesis of  Tunrayo Adeleke-Larodo \cite{Tun} and were also independently  derived by Luca Giuggioli. We thank them for
useful discussions.

MRE would like to acknowledge funding from EPSRC under grant number
EP/J007404/1. SNM thanks the Higgs Centre for hospitality
during the writing of the manuscript. DB acknowledges support from PAPIIT (UNAM) grant IN105015.

{}

\end{document}